\documentclass[runningheads,anonymous=true]{llncs}
\usepackage[misc,geometry]{ifsym}
\usepackage{graphicx}
\usepackage{siunitx}
\usepackage{xcolor}
\usepackage{cite}
\makeatletter
\renewcommand\subsubsection{\@startsection{subsubsection}{3}{\z@}%
                       {-8\p@ \@plus -4\p@ \@minus -4\p@}
                       {-0.5em \@plus -0.22em \@minus -0.1em}%
                       {\normalfont\normalsize\bfseries\boldmath}}
\makeatother
\makeatletter
\newcommand{\mypm}{\mathbin{\mathpalette\@mypm\relax}}
\newcommand{\@mypm}[2]{\ooalign{%
  \raisebox{.1\height}{$#1+$}\cr
  \smash{\raisebox{-.6\height}{$#1-$}}\cr}}
\makeatother

\begin{document}
%


\title{Unsupervised Clustering of Quantitative Imaging Phenotypes using Autoencoder and Gaussian Mixture Model}

\titlerunning{Unsupervised Clustering of Quantitative Imaging Phenotypes}

\author{Jianan Chen 
\inst{1,3}\and  
Laurent Milot\inst{2,3}\and
Helen M. C. Cheung\inst{2,3}\and
Anne L. Martel\inst{1,3}
}
\authorrunning{J.Chen et al.}

\institute{
Department of Medical Biophysics, University of Toronto, Toronto, ON, CA 
\\
\email{anne.martel@sri.utoronto.ca} \and
Department of Medical Imaging, University of Toronto, Toronto, ON, CA \and
Sunnybrook Research Institute, Toronto, ON, CA
}

\maketitle              
\begin{abstract}
Quantitative medical image computing (radiomics) has been widely applied to build prediction models from medical images. However, overfitting is a significant issue in conventional radiomics, where a large number of radiomic features are directly used to train and test models that predict genotypes or clinical outcomes. In order to tackle this problem, we propose an unsupervised learning pipeline composed of an autoencoder for representation learning of radiomic features and a Gaussian mixture model based on minimum message length criterion for clustering. By incorporating probabilistic modeling, disease heterogeneity has been taken into account. The performance of the proposed pipeline was evaluated on an institutional MRI cohort of 108 patients with colorectal cancer liver metastases. Our approach is capable of automatically selecting the optimal number of clusters and assigns patients into clusters (imaging subtypes) with significantly different survival rates. Our method outperforms other unsupervised clustering methods that have been used for radiomics analysis and has comparable performance to a state-of-the-art imaging biomarker. 

\end{abstract}

\section{Introduction}
Quantitative medical image computing, known as radiomics, has been an emerging field in medical image analysis. The goal of radiomics is to extract hundreds of features that are mathematical summarizations of the volume-of-interest (VOI) from computed tomography (CT), magnetic resonance imaging (MRI) or positron emission tomography (PET), for the purpose of disease characterization and patient stratification. Imaging biomarkers built on radiomic features have demonstrated great performance in predicting patients outcome~\cite{ha2017metabolic,Aerts2014,kontos2018radiomic}. However, there are two critical issues in radiomics analysis.

First, there is always the possibility of overfitting. Building statistically powerful models at medically meaningful effect sizes using hundreds of radiomic features would at least require thousands of samples \cite{Napel2018}, while such datasets are rarely available for research in medical settings. Even with generalization techniques such as feature reduction and cross-validation, overfitting remains a primary concern.

Second, conventional radiomics analysis does not necessarily consider the heterogeneity of the disease of interest. Breast cancer imaging subtypes identified using a radiomics approach are distinct from established breast cancer molecular and pathological subtypes \cite{Wu2017}. This suggests that tumors in different subtypes or genotypes may have similar appearances. However, radiomic models are usually trained with either linear or deterministic models such as logistic regression or support vector machines for subtype prediction. These approaches are not ideal for the modeling of overlapping distributions. 

Unsupervised clustering can be leveraged to alleviate these problems. 
Consensus clustering (CC), for example, has been used with radiomic features extracted from the tumor and surrounding parenchyma to define imaging subtypes in breast cancer~\cite{monti2003consensus,Wu2017}. However, it is difficult to glean further insights of the imaging subtypes through clustering without additional analyses, and often non-trivial to select the optimal number of clusters.

In this paper, we alleviated the issues of overfitting and non-distinct feature distributions using a radiomics pipeline that identifies imaging subtypes based on radiomic features using a probabilistic generative model with minimum supervision. We made use of an autoencoder to reduce the dimensions of our radiomic features and find representations with minimal correlations. We incorporated a Gaussian mixture model (GMM) with minimum message length criterion (MML) to cluster patient MRIs into imaging subtypes using the learned representations of features \cite{figueiredo2002unsupervised}. We compared the performance of other clustering algorithms with our approach and investigated the clinical value of the defined imaging subtypes. In addition, we demonstrated that the imaging subtypes derived from our pipeline have comparable prognostic ability to state-of-the-art clinical and imaging biomarkers.

\section{Method}
\begin{figure}
    \centering
    \includegraphics[scale=0.55]{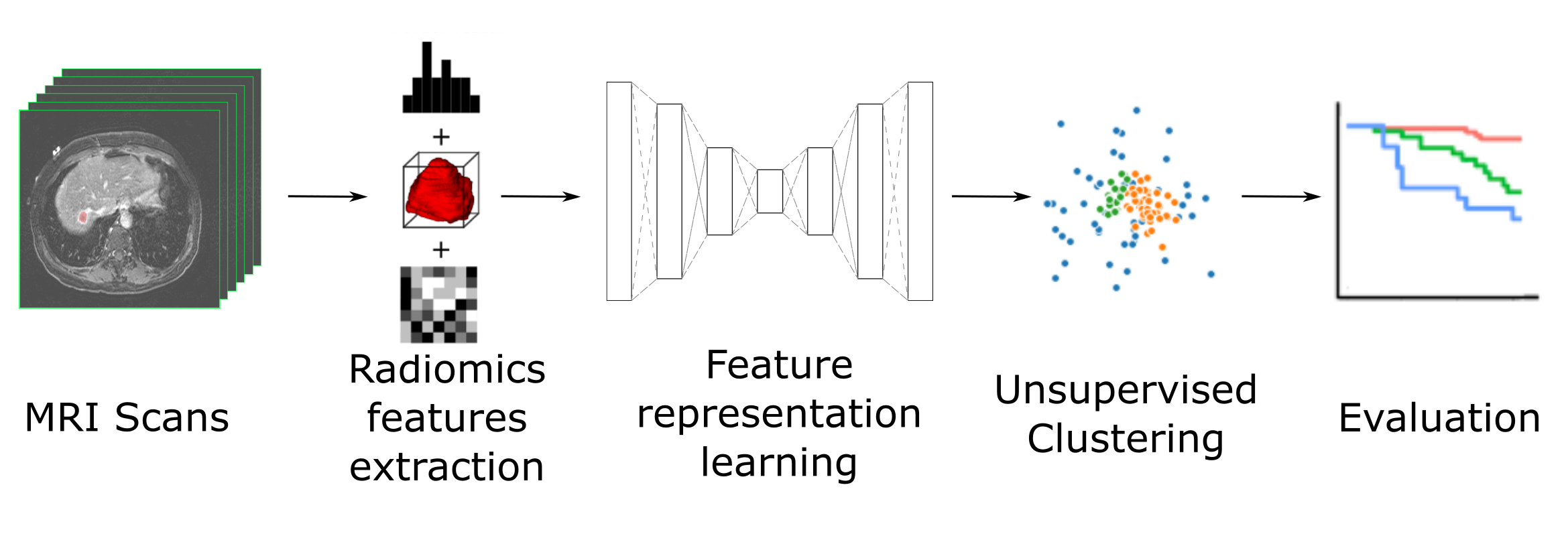}
    \caption{Workflow of the proposed unsupervised radiomics pipeline}
    \label{workflow}
\end{figure}

The workflow of our unsupervised clustering radiomics analysis includes five steps. (Fig. \ref{workflow}). First, quantitative imaging features are extracted from MRI scans. Then an autoencoder is applied to learn feature representations for the purpose of dimension reduction. Next, the learned representations from the latent space are clustered using a GMM. Finally, the significance and clinical value of the learned imaging subtypes (clusters) are evaluated. Clinical outcome is held out during model training and only used in evaluation stage.
\subsubsection{Radiomic Feature Extraction}
The first step of our pipeline was MRI tumour lesion segmentation and radiomic features extraction. The MRIs were resampled to isotropic spacing [3,3,3] using B-spline transformation.
A Z-score transformation was applied to normalize the resampled images and rescale their intensities to range from 0 to 100. Outliers with extreme intensity values were capped in normalization. Image discretization was performed using a bin width of 5 to simplify computation. Finally, 100 features from three categories describing the characteristics (intensity, shape and texture) of the volume of interest, \textit{e.g.} tumour lesions, were extracted using pyradiomics \cite{van2017computational}. 

\subsubsection{Quantile Normalization}
We observed that radiomic features extracted from medical images frequently included outliers, \textit{i.e.} samples with extreme feature values. However, in order to be able to reconstruct features from very few latent features in autoencoders, extreme values should be avoided. Experimentally we found that Z-normalization and capping did not work well for this problem. Hence, we used a quantile normalization approach with thresholds designed for radiomic features. We transform feature quantiles: 0-5\% (min), 5-25\%, 25\%-50\%, 50\%-75\%, 75\%-95\%, 95\%-100\% (max) to floats: 0, 1/6, 2/6, 3/6, 4/6, 5/6, 1, respectively. The quantile thresholds were empirically determined based on experiments. 


\subsubsection{Feature Representation Learning (with autoencoder)}
\begin{figure}
    \centering
    \includegraphics[scale=0.4]{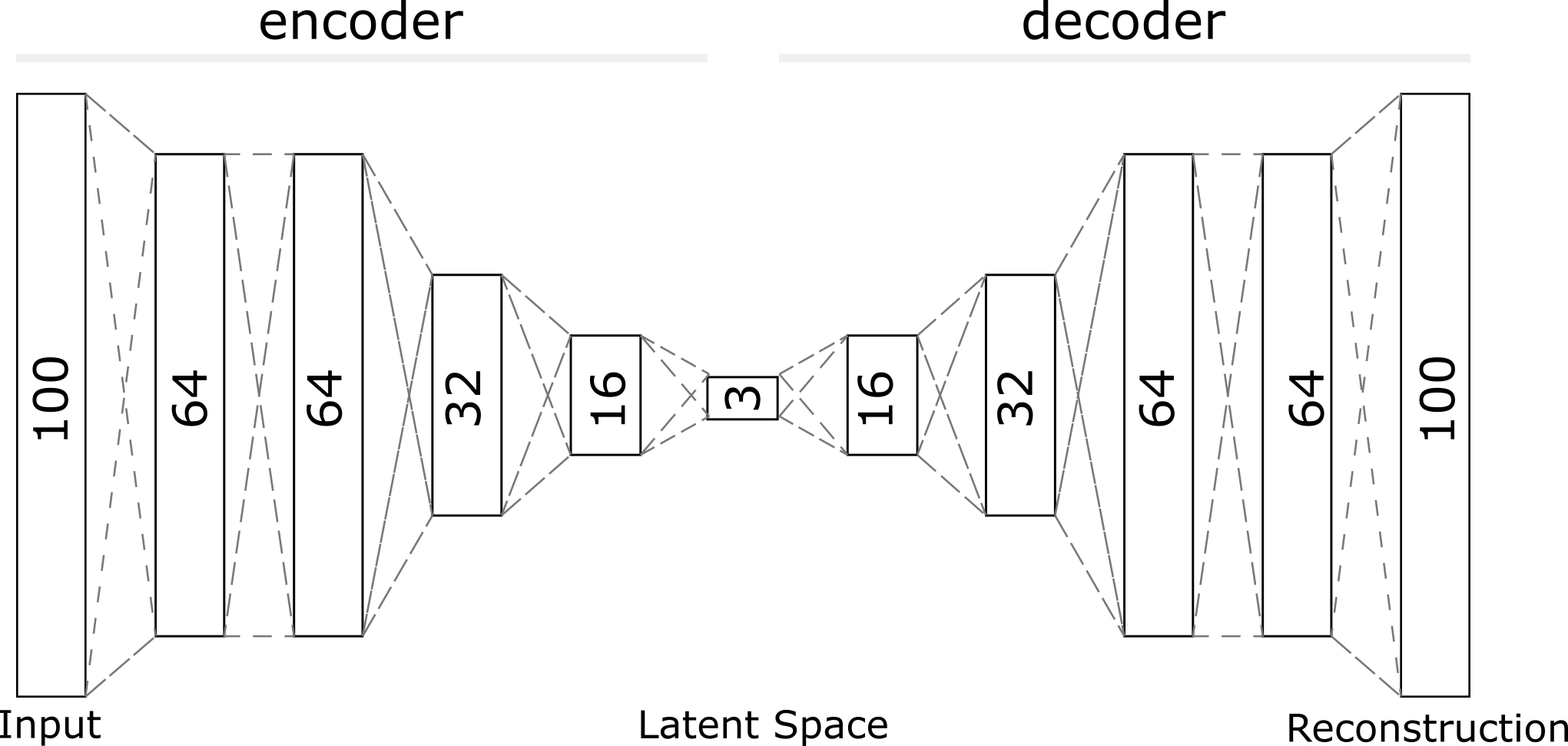}
    \caption{Network architecture of the autoencoder with fully-connected layers used for feature representation learning. The numbers in the blocks denote number of hidden units in each layer.}
    \label{AE}
\end{figure}

With the transformed radiomic features, we leveraged an autoencoder to learn feature representations \textit{i.e.} features in low dimensional latent space that best summarize the data. The input for the autoencoder was the 100-element radiomic feature vector previously mentioned (Fig. \ref{AE}). There were five layers for both the encoder and the decoder. A relatively deep structure compared to the size of input was used to add non-linearity. The autoencoder was blinded to all clinical information to prevent overfitting. All layers were fully connected with SELU activation \cite{klambauer2017self}. We chose three as the number of latent features because there were three categories of radiomic features. The latent features were used as the input for subsequent unsupervised clustering.
\subsubsection{Unsupervised Clustering (with GMM-MML)}
Given the known extant of heterogeneity in cancer, imaging phenotypes of tumors with different genotypes/subtypes cannot be expected to be discrete. This motivated the choice of a probabilistic model instead of a deterministic one for tumor image clustering across known molecular and pathological categories. Therefore, we used a GMM for the unsupervised discovery of imaging subtypes.

Gaussian mixtures are weighted linear combinations of $c$-component Gaussians that are used to model a probability density. The optimum number of components $c$ and parameters $\theta$ are estimated using minimum message length criterion (MML) \cite{figueiredo2002unsupervised}. The idea of MML is to simplify estimation of GMM parameters by minimization of encoding length. Consider a dataset $\mathcal{Y}$ which is generated from a probabilistic distribution $p (\mathcal{Y}|\theta)$, the message length required to encode and transmit $\mathcal{Y}$ is:
\begin{equation}
    \mathcal{L}(\theta,\mathcal{Y}) = \mathcal{L}(\theta)+\mathcal{L}(\mathcal{Y}|\theta))
\end{equation}
where $\theta$ is the prior and $\mathcal{L}$ is the encoding message length.

Then $\theta$ and $c$ can be simultaneously estimated using the following equation:
\begin{equation}
    \hat{\theta} = arg \min_\theta\{-\log p(\theta)- \log p (\mathcal{Y}|\theta) + \frac{1}{2}\log|I(\theta)|+\frac{c}{2}(1+\log\frac{1}{12}) \}
\end{equation}
where $|I(\theta)|$ is the determinant of the expected Fisher information matrix \cite{figueiredo2002unsupervised}.


\section{Experiments and Results}
\subsection{Experimental Design}
\subsubsection{Data Description}
The dataset we used is an institutional retrospective cohort of 108 patients with colorectal cancer liver metastases (CRCLM). The number of slices for each volume is between 57 and 170 (mean=100). Institutional Review Board approved the study and waived informed consent. Gadobutrol-enhanced liver MRIs, specifically 10-minute-delayed T1 MRIs were acquired after chemotherapy and prior to hepatic resection, with 1.5/3T MR systems. Tumor segmentation was performed by a single reader with 6 years of experience. The reader was blinded to all clinical information expect patient history of CRCLM. Patient mortality right-censored at three years were available.
\subsubsection{Implementation Details}
Radiomic feature extraction was implemented using packages pyradiomics (v2.0.0) \cite{van2017computational}, numpy (v1.14.3) and SimpleITK (v1.1.0) in Python (v2.7.15).

An autoencoder was trained with Keras on a NVIDIA TITAN Xp Graphics Card (with 12G memory). The loss function was set to binary cross entropy loss. We used adam optimization with a mini-batch size of 64. Default parameters were used, specifically initial learning rate of 0.001, decay rates of 0.9 and 0.999. The model was trained for 400 epochs. 

GMM with MML was implemented using Python package gmm-mml (v0.11). Initial number of clusters $k\_max$ was set to 25. Maximum iteration of 100 and convergence threshold of \SI{e-5} was used.

\subsubsection{Evaluation Metrics}
In radiomic subtype classification tasks, there is no ``ground truth". Thus, instead of using normalized mutual information and Rand index, which are typically used for evaluating the performances of unsupervised clustering methods, we evaluated the clinical value of our imaging clusters according to their association with patient outcome. Cox proportional hazard models were built for comparing different unsupervised clustering methods, in terms of concordance index (CI), hazard ratio (HR) and p-value. A cluster number of three was automatically selected by both GMM-MML and the heuristic algorithms provided in SIMLR (v1.8.1) \cite{wang2018simlr}, so we compared the methods by the maximum pair-wise HR produced from their clustering results.

\subsubsection{Evaluation}
The resulting clusters from our pipeline and their clinical values were demonstrated using a kaplan-meier plot and results from a log-rank test. We also compared our approach with other unsupervised clustering methods that have been used to cluster radiomic features, including baseline consensus clustering\cite{monti2003consensus,Wu2017} and state-of-the-art SIMLR\cite{wang2018simlr}. Further, we compared our approach to validated clinical and imaging biomarkers for prognostic ability \cite{fong1999clinical,cheung2018late}.
\begin{table}[]
    \centering
    \begin{tabular*}{\textwidth}{l@{\extracolsep{\fill}}lll}
    \hline
        \textbf{Method} & \textbf{Concordance index$\uparrow$} & \textbf{Hazard ratio$\uparrow$} & \textbf{P value} \\ \hline 
        CC\cite{monti2003consensus} & 0.531$\pm$0.046 & 2.90 (0.75-14.45) & 0.113\\ \hline
       SIMLR\cite{wang2018simlr} & 0.582$\pm$0.047 & 3.20 (0.73-13.93) & 0.122\\ \hline
        \textbf{AE+GMM+MML} & \textbf{0.623$\pm$0.052} & \textbf{3.32 (1.35-8.18)} & \textbf{0.009}*\\ \hline
    \end{tabular*}
    \caption{Comparison of unsupervised learning algorithms for imaging phenotype clustering. Numbers in parentheses are confidence intervals.``*" denotes significant p value.}
    \label{comparison}
\end{table}
\subsection{Results}
\begin{figure}
    \centering
    \includegraphics[scale=0.63]{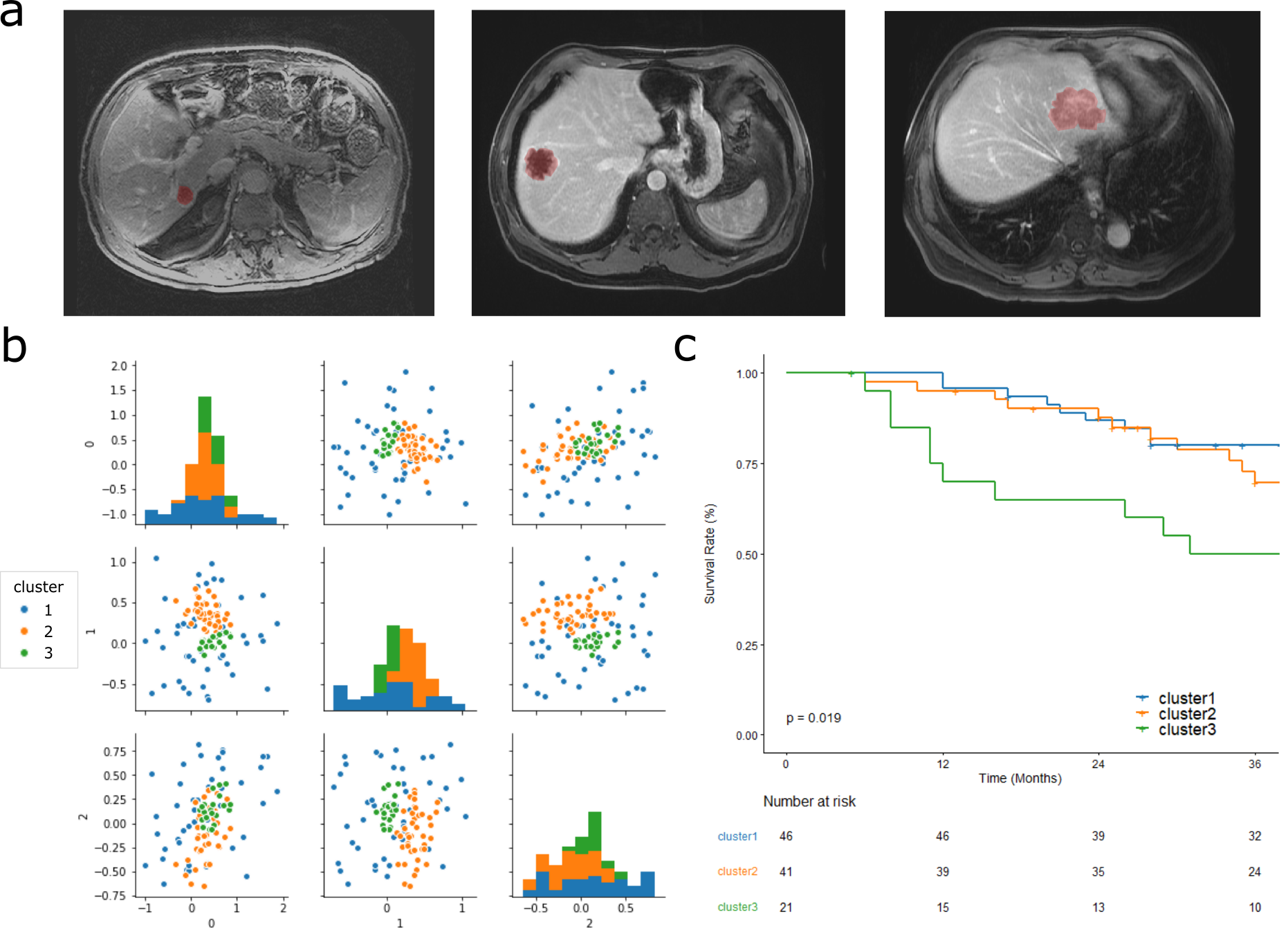}
    \caption{Image subtypes defined by our pipeline (a) from left to right, example images that are representative of cluster I, II and III tumors, respectively. Tumor regions of interest are marked in red. (b) latent features and the estimated Gaussian mixtures. The x- and y-axis of each subplot are latent features values (indexed 0-2). (c) 3-year survival rates for patients assigned to different subtypes.}
    \label{results}
\end{figure}

We estimated Gaussian mixtures of latent representations of radiomic features from liver MRIs using our pipeline. Three Gaussian distributions were learned, each representing an imaging subtype (Figure \ref{results}.a-b). We compared the raw radiomic feature distributions of the three tumor subtypes. We found subtype II tumors to be larger but otherwise very similar to subtype I tumors. In contrast, tumors in subtype III were drastically different from the other two, mostly with higher heterogeneity in texture and were more elongated. There are 46, 41 and 21 patients in cluster I-III, respectively. Patients in subtype III had significantly worse survival rates than patients in the other two subtypes (Figure \ref{results}.c). 



We compared our approach and other unsupervised clustering algorithms for defining imaging subtypes (Table \ref{comparison}). Our approach achieved the highest CI and the only statistically significant HR of 3.32 [1.35-8.18]. Therefore our approach was uniquely able to produce imaging subtypes predictive of patient survival. In addition, the estimated GMM components can be used to model theoretical distributions of CRCLM tumor appearance and be expanded to a validated training-testing model, while the subtypes defined by SIMLR and consensus clustering are hard to interpret and cannot be expanded to incorporate new samples. 

We also compared the prognostic ability of our approach to other biomarkers for CRCLM (Table \ref{tte}). The Fong score is a clinical risk score based on five independent preoperative risk factors built to predict patient prognosis \cite{fong1999clinical}. Target-tumor-enhancement (TTE) is an imaging biomarker specifically designed for stratifying CRCLM patients with late-gadolinium-enhanced MRI \cite{cheung2018late}. The comparison was based on a subset of 99 patients who had all three biomarkers available. Our approach outperformed the Fong score and had comparable performance to TTE in predicting 3-year survival rates for CRCLM patients.

\begin{table}[]
    \centering
    \begin{tabular*}{\textwidth}{l@{\extracolsep{\fill}}ll}
    \hline
        \textbf{Method} & \textbf{Hazard ratio$\uparrow$} & \textbf{P value} \\ \hline 
        Fong score \cite{fong1999clinical}  & 2.28 [0.96-5.42] & 0.060\\ \hline
       \textbf{Target tumor enhancement\cite{cheung2018late}}  & \textbf{4.06 [1.73-9.51]} & \textbf{0.001*}\\ \hline
        \textbf{AE+GMM+MML}  & \textbf{3.98 [1.60-9.89]} & \textbf{0.003*}\\ \hline
    \end{tabular*}
    \caption{Comparison of our approach with clinical and imaging biomarkers for CRCLM. The methods are adjusted for age and sex. Numbers in parentheses are confidence intervals.``*" denotes significant p value.}
    \label{tte}
\end{table}

\section{Discussion and Conclusion}
Probabilistic generative modeling for unsupervised clustering can alleviate the issues of overfitting and overlapping feature distributions in radiomics analysis. We proposed an unsupervised pipeline composed of an autoencoder and a GMM-MML for identifying imaging subtypes from radiomic features that achieved clinically meaningful results. Experiments showed that our approach outperformed other unsupervised algorithms typically used in radiomics by finding imaging subtypes that were associated with patient survival. We also demonstrated that our unsupervised model outperformed a clinical prognostic biomarker and had comparable performance to a state-of-the-art imaging biomarker designed for this specific tumor type and contrast agent.

Due to the modest sample size, we didn't perform a train-test split for feature representation learning and Guassian mixture modeling. However, since the prediction of patient survival rate was independent of radiomic feature encoding and clustering, the survival difference between imaging subtypes were not due to overfitting in this respect. Also, our study was based on manual segmentations of tumor lesions. In future work, we will apply convolutional neural network structures which can extract features directly from MRI scans and remove the need of manual segmentations.

The underlying biological mechanisms of the identified imaging clusters can be elucidated further using pathway enrichment analysis \cite{reimand2019pathway,Wu2017}. Investigations into targetable recurrent mutational or expressional changes in the subtypes may also open the way to guided personalized treatments for colorectal cancer liver metastases patients based on imaging analysis.

\subsubsection{Acknowledgements}
The authors would like to thank The Natural Sciences and Engineering Research Council of Canada (NSERC) for funding, and acknowledge the contribution of Drs. Karanicolas, Law and Coburn in helping to create the patient cohort for this study.
%
%
%
%

\bibliographystyle{splncs04}
\bibliography{paper696}
\end{document}